# *Homogenization of Dispersive Spacetime Crystals: Anomalous Dispersion and Negative Stored Energy*

João C. Serra[1,*], Mário G. Silveirinha[1,†]

[1] University of Lisbon–Instituto Superior Técnico and Instituto de Telecomunicações, Avenida Rovisco Pais, 1, 1049-001 Lisboa, Portugal

**Abstract**

We introduce a homogenization approach to characterize the dynamical response of a generic dispersive spacetime crystal in the long-wavelength limit. The theory is applied to dispersive spacetime platforms with a travelling-wave modulation. It is shown that for long wavelengths the effective response may be described by a frequency dependent permittivity. Due to the active nature of spacetime systems, the permittivity is not bound by the same constraints as in standard time-invariant metamaterials. In particular, we find that dispersive spacetime crystals can exhibit rather peculiar physics, such as an anomalous ("non-Foster") permittivity dispersion with a negative stored energy density, alternate between gain and loss regimes, and present multiple resonances in the quasi-static regime. Furthermore, it is verified with numerical simulations that the effective theory captures faithfully the exact dispersion of the first few photonic bands.

* E-mail: *joao.leandro.camara.serra@tecnico.ulisboa.pt*
† To whom correspondence should be addressed: E-mail: *mario.silveirinha@co.it.pt*

# I. Introduction

Metamaterials have emerged in the last decades as unique platforms to sculpt light-matter interactions [1-2]. Traditionally, metamaterials have been engineered by tailoring in shape and in size identical meta-atoms arranged in a space-lattice. In recent years, time has been explored as a new degree of freedom in material design, typically relying on an electrical modulation of the material parameters [3-11]. Time-varying materials are particularly appealing because they enable non-reciprocal electromagnetic interactions without the need for an external magnetic field bias [12-15].

In the long-wavelength limit, the electrodynamics of metamaterials can be conveniently modeled using homogenization methods [16-28]. Effective medium theory not only greatly simplifies the description of the wave propagation but may also be used to identify the physical mechanisms that control the relevant electromagnetic phenomena. Thus, the effective medium formalism combines the simplicity of the analytical modeling with invaluable physical insights. Although homogenization theories for spacetime metamaterials have been recently proposed in the literature [29-34], they are typically restricted to the quasi-static regime and do not capture the inherent frequency dispersion of the effective material response. It was shown in Ref. [34], that temporal metamaterials (varying periodically in time, but uniform in space) can be characterized by strong spatial dispersion in the long-wavelength limit.

In this work, we introduce an effective medium description of the long-wavelength response of a dispersive spacetime crystal with a travelling-wave modulation. For a travelling-wave modulation the relevant material parameters vary in spacetime as $u(\mathbf{r},t) = u(\mathbf{r} - \mathbf{v}t)$ with $\mathbf{v}$ the modulation speed and $u$ some generic parameter that is controlled by the modulation. Our analysis fully takes into account the inevitable material dispersion of the constituents of the metamaterial and determines the effective frequency-dependent material response of the system. The developed theory may be regarded as a generalization of the source-driven homogenization introduced in Ref. [18, 26] to the case of spacetime crystals. The effect of material dispersion in time-varying systems was studied in previous works in different contexts [35-39].

Notably, we find that the complex interactions resulting from the spacetime modulation may yield an active effective response characterized by an anomalous ("non-Foster") material dispersion and a negative stored energy density. Thus, the spacetime modulation can be used to engineer non-Foster metamaterials. Non-Foster dispersions are instrumental to boost the bandwidth of waveguides, antennas, and related microwave components [40-45].

The paper is organized as follows. Section II extends the formalism of source-driven homogenization theory to spacetime metamaterials. The analysis is focused on crystals with a travelling-wave type modulation. Section III introduces the dispersive spacetime model for the stratified photonic crystal considered in this work. Analytical formulas for the effective permittivity are derived under a quasi-static approximation. Then, in Sect. IV we study different manifestations of the non-Hermitian properties of the spacetime crystal in the effective response. In particular, it is shown that the effective permittivity may present an anomalous (non-Foster) dispersion. Furthermore, due to the active nature of the material response, the energy supplied by an external excitation to reach a steady state may be negative. We also discuss the conditions under which the effective permittivity of the dispersive spacetime crystal may become complex-valued. In section V, the validity of the effective medium theory is demonstrated by comparing the exact band structure with the predictions obtained from the homogenization approach. Additionally, we describe how the combination of dispersion and spacetime band folding results in elaborate band structures. Section VI presents the main conclusions.

## II. Source-driven homogenization of spacetime metamaterials

In this section, we generalize the formalism of source-driven homogenization theory to spacetime metamaterials. The analysis is focused on the homogenization of metamaterials with a travelling-wave modulation.

*A. Source-driven homogenization*

Let us consider Maxwell's equations in a generic time-variant material:

$$-\nabla \times \mathbf{E} = \partial_t \left[ \mu_0 \mathbf{H} + \mathbf{P}_m \right] + \mathbf{j}_{m,\text{ext}}$$
$$+\nabla \times \mathbf{H} = \partial_t \left[ \varepsilon_0 \mathbf{E} + \mathbf{P}_e \right] + \mathbf{j}_{e,\text{ext}}. \quad (1)$$

Here, $\mathbf{P}_e$, $\mathbf{P}_m$ are the electric and magnetic polarization vectors, respectively, which determine the material response through some suitable constitutive relations. Furthermore, $\mathbf{j}_{e,\text{ext}}$, $\mathbf{j}_{m,\text{ext}}$ are fictitious electric and magnetic external excitations that drive the system. We consider both electric and magnetic excitations, as both are relevant to formulate the homogenization problem, as detailed below.

The objective of homogenization theory is to characterize the slowly varying in spacetime envelopes of the electromagnetic fields $\langle \mathbf{E} \rangle$ and $\langle \mathbf{H} \rangle$. Here, $\langle \mathbf{\Psi} \rangle$ represents a "spacetime-averaged" field $\mathbf{\Psi}$ defined by a spacetime convolution:

$$\langle\Psi\rangle(\mathbf{r},t) = \iint d^3\mathbf{r}_0 dt_0\, f(\mathbf{r}-\mathbf{r}_0, t-t_0)\Psi(\mathbf{r}_0, t_0). \tag{2}$$

Analogous to Refs. [16-18, 25-26], it is assumed that the kernel $f$ corresponds to an ideal low-pass spacetime filter such that:

$$\langle e^{-i\omega t}e^{i\mathbf{k}\cdot\mathbf{r}}\rangle(\mathbf{r},t) = e^{-i\omega t}e^{i\mathbf{k}\cdot\mathbf{r}}\tilde{f}(\mathbf{k},\omega), \quad \text{with} \quad \tilde{f}(\mathbf{k},\omega) = \begin{cases} 1, & (\mathbf{k},\omega)\in\text{BZ} \\ 0, & \text{otherwise} \end{cases}. \tag{3}$$

In the above, $\tilde{f}(\mathbf{k},\omega) = \iint d^3\mathbf{r}dt\, f(\mathbf{r},t)e^{i\omega t}e^{-i\mathbf{k}\cdot\mathbf{r}}$ is the Fourier transform of $f$ and BZ is some subset of the spectral domain, which will be taken as a Brillouin zone. When the system under study is time invariant, the time averaging is unnecessary and one may take $f(\mathbf{r}-\mathbf{r}_0, t-t_0) = f_0(\mathbf{r}-\mathbf{r}_0)\delta(t-t_0)$ [16-18, 25-26].

In this article, it is supposed that the system is periodic both in space and in time, such that the constitutive relations are described by some operator $\hat{L}$ with the same property: $\hat{L}(\mathbf{r},t,i\partial_t) = \hat{L}(\mathbf{r}+\mathbf{a}_i,t,i\partial_t)$ ($i$=1,2,3) and $\hat{L}(\mathbf{r},t,i\partial_t) = \hat{L}(\mathbf{r},t+T,i\partial_t)$. Here $\mathbf{a}_i$ are the spatial-lattice primitive vectors, and $T$ is the time-period. Thereby, the set BZ may be taken as the Cartesian product of the space and time Brillouin zones. For example, for a 1D spacetime crystal one may take $\text{BZ} = \left[\frac{-\pi}{a},\frac{\pi}{a}\right]\times\left[\frac{-\pi}{T},\frac{\pi}{T}\right]$.

The constitutive relations are of the generic form:

$$\hat{L}(\mathbf{r},t,i\partial_t)\cdot\begin{pmatrix}\mathbf{P}_e \\ \mathbf{P}_m\end{pmatrix} = \begin{pmatrix}\varepsilon_0\mathbf{E} \\ \mu_0\mathbf{H}\end{pmatrix}. \tag{4}$$

Thus, the dynamics of the electric and magnetic polarization vectors is controlled by a time differential equation whose coefficients may vary in space and in time.

The goal of the source-driven homogenization is to predict the dynamics of the averaged fields $\langle \mathbf{E} \rangle$ and $\langle \mathbf{H} \rangle$ for any macroscopic excitation $\mathbf{j} = \begin{pmatrix} \mathbf{j}_{e,ext} & \mathbf{j}_{m,ext} \end{pmatrix}^T$ that varies slowly in spacetime, such that $\langle \mathbf{j} \rangle = \mathbf{j}$ [17, 26, 28]. As both the space and time derivatives commute with the averaging operator, the averaged fields satisfy:

$$\begin{aligned} -\nabla \times \langle \mathbf{E} \rangle &= \partial_t \left[ \mu_0 \langle \mathbf{H} \rangle + \langle \mathbf{P}_m \rangle \right] + \mathbf{j}_{m,ext} \\ +\nabla \times \langle \mathbf{H} \rangle &= \partial_t \left[ \varepsilon_0 \langle \mathbf{E} \rangle + \langle \mathbf{P}_e \rangle \right] + \mathbf{j}_{e,ext} \end{aligned}. \quad (5)$$

We would like to relate the averaged polarization vectors $\langle \mathbf{P}_e \rangle$, $\langle \mathbf{P}_m \rangle$ with the macroscopic fields $\langle \mathbf{E} \rangle$, $\langle \mathbf{H} \rangle$ through some linear operator $\hat{\mathbf{M}}_{ef}$ that describes the response of the effective medium:

$$\begin{pmatrix} \langle \mathbf{P}_e \rangle \\ \langle \mathbf{P}_m \rangle \end{pmatrix} = \hat{\mathbf{M}}_{ef} \cdot \begin{pmatrix} \langle \mathbf{E} \rangle \\ \langle \mathbf{H} \rangle \end{pmatrix} - \begin{pmatrix} \varepsilon_0 \langle \mathbf{E} \rangle \\ \mu_0 \langle \mathbf{H} \rangle \end{pmatrix}. \quad (6)$$

Note that the above equation is equivalent to $\begin{pmatrix} \langle \mathbf{D} \rangle \\ \langle \mathbf{B} \rangle \end{pmatrix} = \hat{\mathbf{M}}_{ef} \cdot \begin{pmatrix} \langle \mathbf{E} \rangle \\ \langle \mathbf{H} \rangle \end{pmatrix}$, with $\mathbf{D}$ and $\mathbf{B}$ the electric displacement vector and the induction vector, respectively.

Evidently, external excitations that satisfy $\langle \mathbf{j} \rangle = \mathbf{j}$ are of the form $\mathbf{j} = \sum_{(\mathbf{k},\omega) \in \mathrm{BZ}} \mathbf{j}^0_{(\mathbf{k},\omega)} e^{-i\omega t} e^{i \mathbf{k} \cdot \mathbf{r}}$ with $\mathbf{j}^0_{(\mathbf{k},\omega)}$ some constant "vector". The key point is that, since the problem is linear, the unknown (linear) operator $\hat{\mathbf{M}}_{ef}$ is fully determined by the response of the system to spacetime-harmonic excitations of the type $\mathbf{j}_0 e^{-i\omega t} e^{i \mathbf{k} \cdot \mathbf{r}}$ with $(\mathbf{k}, \omega) \in \mathrm{BZ}$ [17, 26, 28]. Therefore, similar to the case of time-invariant metamaterials, it is possible to determine $\hat{\mathbf{M}}_{ef}$ in the spectral domain ($\mathbf{M}_{ef} = \mathbf{M}_{ef}(\mathbf{k}, \omega)$) by finding the solution of Eq. (1) for a generic excitation of the type $\mathbf{j}_0 e^{-i\omega t} e^{i\mathbf{k}\cdot\mathbf{r}}$ followed by field averaging [Eq. (2)]. It should be noted that for an excitation of the form $\mathbf{j}_0 e^{-i\omega t} e^{i\mathbf{k}\cdot\mathbf{r}}$ the solution of Eq. (1) is such that a generic field is of the form $\mathbf{\Psi}(\mathbf{r},t) = \mathbf{\Psi}_p(\mathbf{r},t) e^{-i\omega t} e^{i\mathbf{k}\cdot\mathbf{r}}$ with $\mathbf{\Psi}_p(\mathbf{r},t)$ a periodic function of spacetime. Using this property, it can be checked that:

$$\langle \mathbf{\Psi} \rangle (\mathbf{r},t) = e^{-i\omega t} e^{+i\mathbf{k}\cdot\mathbf{r}} \frac{1}{T} \frac{1}{V_{cell}} \int_{cell} d^3\mathbf{r} \int_0^T dt\, \mathbf{\Psi}(\mathbf{r},t) e^{i\omega t} e^{-i\mathbf{k}\cdot\mathbf{r}}, \tag{7}$$

where $V_{cell}$ is the volume of the unit cell. The quantities $\langle \mathbf{P}_e \rangle$, $\langle \mathbf{P}_m \rangle$ are determined by similar formulas.

*B.    Travelling-wave modulations*

For generic spacetime variations, the solution of Eq. (1) for an excitation of the type $\mathbf{j}_0 e^{-i\omega t} e^{i\mathbf{k}\cdot\mathbf{r}}$ requires the use of time-domain numerical methods (see Ref. [25]). Consider however the particular subclass of spacetime modulations, such that the operator $\hat{L}$ in Eq. (4) is of the form

$$\hat{L} = \hat{L}(\mathbf{r} - \mathbf{v}t, \partial_t), \tag{8}$$

with $\mathbf{v}$ the modulation speed. We shall refer to this type of modulations as a "travelling-wave modulation". In this case, it is possible to get rid of the time dependence of the operator with a simple Galilean coordinate transformation: $\mathbf{r}' = \mathbf{r} - \mathbf{v}t$, $t' = t$ [9, 32]. Noting that $\nabla = \nabla'$ and $\frac{\partial}{\partial t} = \frac{\partial}{\partial t'} - \mathbf{v}\cdot\nabla'$, it follows that the constitutive relations in the primed coordinates are of the form:

$$\hat{L}\left(\mathbf{r}', \frac{\partial}{\partial t'} - \mathbf{v}\cdot\nabla'\right) \cdot \begin{pmatrix} \mathbf{P}_e \\ \mathbf{P}_m \end{pmatrix} = \begin{pmatrix} \varepsilon_0 \mathbf{E} \\ \mu_0 \mathbf{H} \end{pmatrix}. \tag{9a}$$

As seen, the operator $\hat{L}$ becomes time independent in the co-moving frame, which greatly simplifies the homogenization problem. On the other hand, Maxwell's equations [Eq. (1) with $\mathbf{j} = \mathbf{j}_0 e^{-i\omega t} e^{i\mathbf{k}\cdot\mathbf{r}}$] become:

$$-\nabla' \times \mathbf{E} = \left(\frac{\partial}{\partial t'} - \mathbf{v}\cdot\nabla'\right)[\mu_0 \mathbf{H} + \mathbf{P}_m] + \mathbf{j}_{m,0} e^{-i\omega' t'} e^{i\mathbf{k}'\cdot\mathbf{r}'}$$
$$+\nabla' \times \mathbf{H} = \left(\frac{\partial}{\partial t'} - \mathbf{v}\cdot\nabla'\right)[\varepsilon_0 \mathbf{E} + \mathbf{P}_e] + \mathbf{j}_{e,0} e^{-i\omega' t'} e^{i\mathbf{k}'\cdot\mathbf{r}'}$$ (9b)

where $\mathbf{k}' = \mathbf{k}$ and $\omega' = \omega - \mathbf{k}\cdot\mathbf{v}$. It is possible to recover the usual structure of Maxwell's equations with some simple manipulations:

$$-\nabla' \times [\mathbf{E} + \mathbf{v} \times (\mu_0 \mathbf{H} + \mathbf{P}_m)] = \frac{\partial}{\partial t'}[\mu_0 \mathbf{H} + \mathbf{P}_m] + \mathbf{j}'_{m,0} e^{-i\omega' t'} e^{i\mathbf{k}'\cdot\mathbf{r}'}$$
$$+\nabla' \times [\mathbf{H} - \mathbf{v} \times (\varepsilon_0 \mathbf{E} + \mathbf{P}_e)] = \frac{\partial}{\partial t'}[\varepsilon_0 \mathbf{E} + \mathbf{P}_e] + \mathbf{j}'_{e,0} e^{-i\omega' t'} e^{i\mathbf{k}'\cdot\mathbf{r}'}$$ (9b')

The fields $\mathbf{E}' \equiv \mathbf{E} + \mathbf{v} \times (\mu_0 \mathbf{H} + \mathbf{P}_m)$ and $\mathbf{H}' = \mathbf{H} - \mathbf{v} \times (\varepsilon_0 \mathbf{E} + \mathbf{P}_e)$ may be regarded as the transformed fields in the (Galilean) co-moving frame. In the above, $\mathbf{j}'_{e,0} = \mathbf{j}_{e,0} - \mathbf{v}\rho_{e,0}$ and $\mathbf{j}'_{m,0} = \mathbf{j}_{m,0} - \mathbf{v}\rho_{m,0}$ are the equivalent vector current amplitudes in the co-moving frame, with $\rho_{e,0} = \frac{\mathbf{k}}{\omega}\cdot\mathbf{j}_{e,0}$ and $\rho_{m,0} = \frac{\mathbf{k}}{\omega}\cdot\mathbf{j}_{m,0}$ the electric and magnetic charge densities associated with the external currents. Since $\mathbf{j}'_{e,0}$ and $\mathbf{j}'_{m,0}$ are constant vectors, the system (9) may be regarded as a homogenization problem in a time-invariant system, somewhat analogous to the problems studied in Refs. [16-18, 25-26].

The effective medium response $\mathbf{M}_{ef} = \mathbf{M}_{ef}(\mathbf{k}, \omega)$ can be found by solving Eqs. (9a) and (9b') for a generic excitation of the type $\mathbf{j}'_{e,ext} = \mathbf{j}'_{e,0} e^{-i\omega' t'} e^{i\mathbf{k}'\cdot\mathbf{r}'}$ and $\mathbf{j}'_{m,ext} = \mathbf{j}'_{m,0} e^{-i\omega' t'} e^{i\mathbf{k}'\cdot\mathbf{r}'}$. There are six independent macroscopic excitations, with three degrees of freedom purely electric and another three degrees of freedom magnetic. For each of the excitations, the macroscopic fields are found by averaging the corresponding microscopic fields [Eq. (7)]. In the co-moving frame coordinates, a generic field is of the form $\Psi = \Psi_p(\mathbf{r}')e^{-i\omega't'}e^{i\mathbf{k}'\cdot\mathbf{r}'}$ with $\Psi_p(\mathbf{r}')$ a periodic function of space. Hence, the macroscopic fields can be easily evaluated in the co-moving frame coordinates with a simple spatial averaging:

$$\langle\Psi\rangle(\mathbf{r},t) = e^{-i\omega t}e^{+i\mathbf{k}\cdot\mathbf{r}}\frac{1}{V_{cell}}\int_{cell}d^3\mathbf{r}'\,\Psi(\mathbf{r}',t')e^{i\omega't'}e^{-i\mathbf{k}'\cdot\mathbf{r}'}. \tag{10}$$

After $\langle\mathbf{P}_e\rangle$, $\langle\mathbf{P}_m\rangle$, $\langle\mathbf{E}\rangle$, $\langle\mathbf{H}\rangle$ are determined for the six independent excitations, one can find the (6×6) effective material matrix $\mathbf{M}_{ef} = \mathbf{M}_{ef}(\mathbf{k},\omega)$ using

$$\begin{pmatrix}\langle\mathbf{P}_e\rangle\\\langle\mathbf{P}_m\rangle\end{pmatrix} = \mathbf{M}_{ef}(\mathbf{k},\omega)\cdot\begin{pmatrix}\langle\mathbf{E}\rangle\\\langle\mathbf{H}\rangle\end{pmatrix} - \begin{pmatrix}\varepsilon_0\langle\mathbf{E}\rangle\\\mu_0\langle\mathbf{H}\rangle\end{pmatrix}. \tag{11}$$

The dependence of $\mathbf{M}_{ef}$ on $(\mathbf{k},\omega)$ corresponds to a spatially and frequency dispersive response.

To conclude this section, we note that 2D and 3D periodic space crystals subject to a travelling-wave modulation may not yield a system strictly periodic in time, but rather a quasi-crystal type periodicity in time [46]. In fact, consider some function (e.g., the permittivity) that is periodic in space in the co-moving frame so that $u(x',y') = u(x'+a,y') = u(x',y'+a)$ with $a$ the lattice period. Then, the corresponding function evaluated in the laboratory frame, $u(\mathbf{r},t) = u(\mathbf{r}-\mathbf{v}t)$, with $\mathbf{v} = v_x\hat{\mathbf{x}} + v_y\hat{\mathbf{y}}$ the modulation speed, is periodic in time only when $v_x/v_y$ is a rational number.

## III. Homogenization of dispersive 1D spacetime crystals

### A. Model of the spacetime crystal

In order to illustrate the theory developed in the previous section, we consider a 1D non-magnetic photonic crystal ($\mathbf{P}_m = 0$). Without loss of generality, it is supposed that $\mathbf{E} = E(x,t)\hat{\mathbf{y}}$ and $\mathbf{H} = H(x,t)\hat{\mathbf{z}}$. Furthermore, the electric polarization vector in the material is of the form $\mathbf{P}_e = P(x,t)\hat{\mathbf{y}}$. The dynamics of the electromagnetic fields in the laboratory frame is ruled by:

$$-\partial_x E = \mu_0 \partial_t H, \qquad -\partial_x H = \partial_t [\varepsilon_0 E + P] + j_{e,\text{ext}}. \tag{12}$$

We discard the magnetic excitation ($j_{m,\text{ext}} = 0$) because $\mathbf{P}_m = 0$. Indeed, in the absence of a magnetic response at the microscopic level, it is possible to characterize the effective medium using only an equivalent (scalar) permittivity $\varepsilon_{\text{ef}}(k,\omega)$ that relates the mean electric polarization with the mean electric field: $\langle P \rangle = [\varepsilon_{\text{ef}}(k,\omega) - \varepsilon_0] \cdot \langle E \rangle$ [16-18]. For a 1D problem the effective permittivity $\varepsilon_{\text{ef}}(k,\omega)$ can be calculated using a single electric excitation controlled by $j_{e,\text{ext}}$.

In this study, it is supposed that the dynamics of the polarization vector is controlled by a dispersive Drude-Lorentz model with time-varying coefficients (the modulation speed is $\mathbf{v} = v\hat{\mathbf{x}}$):

$$\left[\frac{\partial^2}{\partial t^2} + \omega_0^2(x-vt)\right] P(x,t) = \varepsilon_0 \omega_p^2(x-vt) E(x,t). \tag{13}$$

Related models have been considered by other authors [47, 48]. In the above, $\omega_0$ is the resonance frequency of the Drude-Lorentz model and $\omega_p$ is the plasma frequency. The two frequencies may be subject to a spacetime travelling-wave modulation with modulation speed $v$. In the absence of a time modulation ($v=0$), the polarization $P$ in the frequency domain is linked to the electric field as: $P_\omega = \frac{\varepsilon_0 \omega_p^2}{\omega_0^2 - \omega^2} E_\omega$. Note that $\omega_0$ typically describes the interaction of the electromagnetic field with bound charges, and so it may be dynamically modulated by applying a strong time-varying electric bias to a nonlinear dielectric material (similar to the electric bias acting on a "varactor"). On the other hand, the parameter $\omega_p$ is mainly controlled by the number of charges in the system and hence its modulation may require charge injection/removal (e.g., using a gate, analogous to a transistor).

Comparing Eqs. (4) and (13), we see that the operator $\hat{L}$ is given by (from now on we ignore the magnetic degrees of freedom):

$$\hat{L}(x,t,\partial_t) = \frac{1}{\omega_p^2(x-vt)} \left[ \frac{\partial^2}{\partial t^2} + \omega_0^2(x-vt) \right]. \tag{14}$$

Switching to the co-moving frame coordinates $(x',t') = (x-vt, t)$, one obtains a simplified version of the system (9):

$$-\frac{\partial}{\partial x'}(E - v\mu_0 H) = \mu_0 \frac{\partial}{\partial t'} H,$$
$$-\frac{\partial}{\partial x'}(H - v\varepsilon_0 E - vP) = \frac{\partial}{\partial t'}[\varepsilon_0 E + P] + j_{e,0} e^{-i\omega' t'} e^{ik'x'} \tag{15a}$$

$$\frac{1}{\omega_p^2(x')}\left[\left(\frac{\partial}{\partial t'} - v\frac{\partial}{\partial x'}\right)^2 + \omega_0^2(x')\right]P = \varepsilon_0 E. \tag{15b}$$

with $k' = k$ and $\omega' = \omega - vk$. Note that Eq. (15a) is the counterpart of (9b') as in the 1D problem $j'_{e,0} = j_{e,0}$.

By definition, the effective permittivity must satisfy $\varepsilon_{ef}(k,\omega) = \varepsilon_0 + \langle P \rangle / \langle E \rangle$, where $P$ and $E$ are the (source-driven) solutions of Eq. (15). In the next subsection, we obtain a quasi-static solution of the homogenization problem.

*B.    Quasi-static limit*

To begin with, we note that as the coefficients of the differential system (15) are time-independent, one can use $\frac{\partial}{\partial t'} = -i\omega'$. Here, we want to focus on the quasi-static limit such that the fields vary slowly both in space and in time and the excitation is periodic in space ($k \approx 0$). As the relevant length scale in the co-moving frame is the lattice period $a$, the condition $k \approx 0$ should be understood as equivalent to $ka \ll 1$. This long wavelength condition is unchanged in the lab frame coordinates because the Galilean transformation does not contract distances.

In the discussed quasi-static regime, the left-hand sides and the right-hand sides of the two equations in (15a) vanish independently. In particular, this means that $\frac{\partial}{\partial x'}E' \approx 0$ and $\frac{\partial}{\partial x'}H' \approx 0$ with the primed fields defined by:

$$\begin{aligned}E' &= E - v\mu_0 H \\ H' &= H - v(\varepsilon_0 E + P)\end{aligned}. \tag{16}$$

Thus, the primed fields can be assumed constant in the quasi-static limit. This approximation is mostly accurate when $v^2/c^2 \ll 1$. It is relevant to note that since we take $k \approx 0$, there is no Doppler shift, so that $\omega' \approx \omega$.

As the original fields can be written in terms of the primed fields as

$$E = \frac{E' + \mu_0 v H' + \mu_0 v^2 P}{1 - v^2/c^2}, \qquad H = \frac{H' + \varepsilon_0 v E' + v P}{1 - v^2/c^2}, \tag{17}$$

it follows from Eq. (15b) that the polarization vector satisfies:

$$\hat{L}_{co} P(x') = \varepsilon_0 \frac{E' + \mu_0 v H'}{1 - v^2/c^2}, \quad \text{with} \tag{18a}$$

$$\hat{L}_{co} P(x') \equiv \left[ \frac{-1}{\omega_p^2(x')} \left( -\omega + v i \frac{\partial}{\partial x'} \right)^2 + \frac{\omega_0^2(x')}{\omega_p^2(x')} - \frac{v^2/c^2}{1 - v^2/c^2} \right] P(x'). \tag{18b}$$

Since the excitation has $k \approx 0$, the polarization $P(x')$ is determined by the periodic solution of the above equation. Thus, the quasi-static solution neglects the effects of spatial dispersion so that the effective permittivity will depend exclusively on frequency. Interestingly, the excitation current $j_{e,0}$ does not appear explicitly in Eq. (18a), but only indirectly through the primed fields $E'$, $H'$.

Let $G(x')$ be the periodic solution of the equation $\hat{L}_{co} G(x') = 1$ with $\hat{L}_{co}$ defined as in Eq. (18b). As $E'$, $H'$ are approximately constants, one can write $P(x') = G(x') \varepsilon_0 \frac{E' + \mu_0 v H'}{1 - v^2/c^2}$. For $k = 0$, the macroscopic polarization vector is given by $\langle P \rangle = P_{av}(\omega) = \frac{1}{a} \int_0^a P(x') dx'$. This shows that:

$$\langle P \rangle = \varepsilon_0 \chi'_{\text{ef}}(\omega) \frac{E' + \mu_0 v H'}{1 - v^2/c^2} \qquad \text{with} \qquad \chi'_{\text{ef}}(\omega) = \frac{1}{a} \int_0^a G(x') dx'. \tag{19}$$

Recalling that $E(x') = \frac{E' + \mu_0 v H'}{1 - v^2/c^2} + \frac{\mu_0 v^2}{1 - v^2/c^2} P(x')$, it is clear that

$\langle E \rangle = \frac{E' + \mu_0 v H'}{1 - v^2/c^2} + \frac{\mu_0 v^2}{1 - v^2/c^2} \langle P \rangle$. Combining this result with Eq. (19), one finds that the effective relative permittivity in the laboratory (unprimed) frame in the quasi-static limit ($\varepsilon_{\text{ef}}(\omega) \equiv \varepsilon_{\text{ef}}(k \approx 0, \omega)$) is given by:

$$\varepsilon_{\text{ef}}(\omega) = 1 + \frac{\langle P \rangle}{\varepsilon_0 \langle E \rangle} = 1 + \frac{\chi'_{\text{ef}}(\omega)}{1 + \frac{v^2/c^2}{1 - v^2/c^2} \chi'_{\text{ef}}(\omega)}. \tag{20}$$

In summary, the solution of $\hat{L}_{\text{co}} G(x') = 1$ determines the effective response of the spacetime crystal in the long-wavelength limit $ka \ll 1$. In subsection III.D, we derive an explicit analytical formula for the case of a binary crystal.

C.    *Symmetries of $\varepsilon_{\text{ef}}(\omega)$*

There are a few symmetries that guarantee that the effective permittivity of the spacetime crystal is real-valued for real-valued frequencies. Specifically, it is demonstrated in Appendix A that when the operator $\hat{L}_{\text{co}} = \hat{L}_{\text{co}}(x', i\partial_{x'}, \omega)$ defined by Eq. (18b) is either Hermitian with respect to the canonical inner product, $\hat{L}_{\text{co}} = \hat{L}_{\text{co}}^\dagger$, or, alternatively, parity-time (PT) symmetric [49-52] such that $\hat{L}_{\text{co}}(x', i\partial_{x'}, \omega) = \hat{L}_{\text{co}}(-x', -i\partial_{x'}, \omega)^*$, then $\varepsilon_{\text{ef}}(\omega)$ is forcibly real-valued.

It is evident from physical considerations that the homogenized system must inherit the same symmetries as the microscopic spacetime crystal. A uniform (space

-14-

independent) dielectric exhibits PT symmetry if and only if $\varepsilon_{\text{ef}}(\omega) = \varepsilon_{\text{ef}}^*(\omega^*)$. Thus, it follows that the effective permittivity of an arbitrary PT-symmetric crystal must be real-valued, which aligns with our theoretical framework.

### D. Binary crystal

Consider now that the spacetime crystal is formed by two homogeneous layers of thickness $a_n$, $n=1,2$, with each layer described by some parameters $\omega_{p,n}$ and $\omega_{0,n}$ as represented in Fig. 1. The lattice period is $a = a_1 + a_2$.

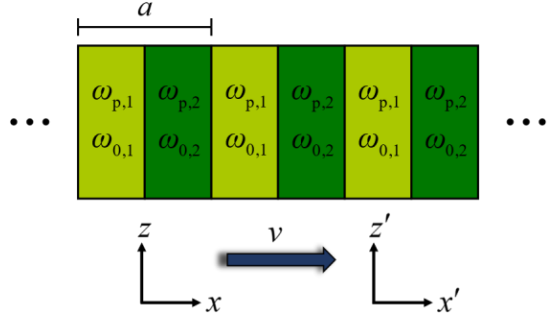

**Fig. 1** Geometry of a one-dimensional binary crystal described by a time-varying Drude-Lorentz model with a travelling-wave modulation.

The solution of Eq. (18) for the binary crystal is of the form:

$$P(x') = \varepsilon_0 \frac{E' + \mu_0 v H'}{1 - v^2/c^2} \times \begin{cases} A_1 e^{i\alpha_1\left(x' - \frac{a_1}{2}\right)} + B_1 e^{i\beta_1\left(x' - \frac{a_1}{2}\right)} + C_1, & 0 \leq x' < a_1 \\ A_2 e^{i\alpha_2\left(x' - \frac{a_2}{2} - a_1\right)} + B_2 e^{i\beta_2\left(x' - \frac{a_2}{2} - a_1\right)} + C_2, & a_1 \leq x' < a \end{cases} \quad (21)$$

with $\quad C_n = \omega_{p,n}^2 \left( \omega_{0,n}^2 - \omega^2 - \omega_{p,n}^2 \frac{v^2/c^2}{1 - v^2/c^2} \right)^{-1}$, $\quad \alpha_n = -\frac{\omega}{v} - \sqrt{\frac{\omega_{0,n}^2}{v^2} - \frac{\omega_{p,n}^2}{v^2} \frac{v^2/c^2}{1 - v^2/c^2}} \quad$ and

$\beta_n = -\frac{\omega}{v} + \sqrt{\frac{\omega_{0,n}^2}{v^2} - \frac{\omega_{p,n}^2}{v^2} \frac{v^2/c^2}{1 - v^2/c^2}}$, $n=1,2$.

The unknown coefficients $A_n$, $B_n$ are determined by imposing the continuity of $P$ and $\partial P / \partial x'$ at the material interfaces. As previously mentioned, in the long-wavelength limit ($ka \ll 1$), the electric polarization $P(x')$ may be assumed periodic in $x'$, and so the relevant boundary conditions are $P(a_1^-) = P(a_1^+)$, $\frac{\partial P}{\partial x'}(a_1^-) = \frac{\partial P}{\partial x'}(a_1^+)$, $P(0^+) = P(a^-)$ and $\frac{\partial P}{\partial x'}(0^+) = \frac{\partial P}{\partial x'}(a^-)$. Explicit formulas for $A_1$, $A_2$, $B_1$, $B_2$ for the case of binary crystals with layers of identical thickness ($a_1 = a_2 = a/2$) are given in Appendix B.

The effective permittivity in the laboratory frame [Eq. (20)] is written in terms of

$$\chi'_{ef}(\omega) = \frac{1}{a}\int_0^a dx' P(x') \left(\varepsilon_0 \frac{E' + \mu_0 v H'}{1 - v^2/c^2}\right)^{-1}$$ [Eq. (19)]. Straightforward calculations show

that for $a_1 = a_2 = a/2$:

$$\chi'_{ef}(\omega) = 2\frac{A_1}{\alpha_1 a}\sin\left(\frac{\alpha_1 a}{4}\right) + 2\frac{B_1}{\beta_1 a}\sin\left(\frac{\beta_1 a}{4}\right) + 2\frac{A_2}{\alpha_2 a}\sin\left(\frac{\alpha_2 a}{4}\right) + 2\frac{B_2}{\beta_2 a}\sin\left(\frac{\beta_2 a}{4}\right) + \frac{C_1 + C_2}{2}.$$

(22)

## IV. Non-Hermitian effects in the effective response of spacetime crystals

### A. *Anomalous permittivity dispersion*

In order to illustrate how the spacetime modulation can tailor in unique ways the effective response, we consider in the first example a binary dispersive crystal with parameters $\omega_{p,1} = \omega_{p,2} = 2.0 c/a$, $\omega_{0,1} = 1.0 c/a$, $\omega_{0,2} = 1.5 c/a$, and $a_1 = a_2 = a/2$. Figures 2ai) and 2bi) depict the effective permittivity of the spacetime crystal (in the laboratory frame) as a function of frequency for the cases $v = 0$ and $v = 0.25c$ (blue lines). The results were calculated using the quasi-static approximation discussed in the previous subsections. The effective permittivity is always real-valued for binary crystals (for the material model in Eq. (18b)), as it can be easily checked that such systems are always PT-invariant. The dashed red lines in Fig. 2 were calculated using a non-dispersive homogenization theory that will be discussed a few paragraphs below.

When $v = 0$ the effective response has two poles coincident with the poles $\omega_{0,n}$ of the individual material responses. In fact, in the static case, the effective permittivity reduces to the spatial average of the material permittivities: $\varepsilon_{\text{ef}}(\omega)\big|_{v=0} = f_1 \varepsilon_1(\omega) + f_2 \varepsilon_2(\omega)$ with $f_n = a_n / a$ the volume fraction of the $n$-th layer and

$$\varepsilon_n(\omega) = 1 + \frac{\omega_{\text{p},n}^2}{\omega_{0,n}^2 - \omega^2}, \tag{23}$$

the dispersive permittivity of the $n$-th material in the absence of the time modulation.

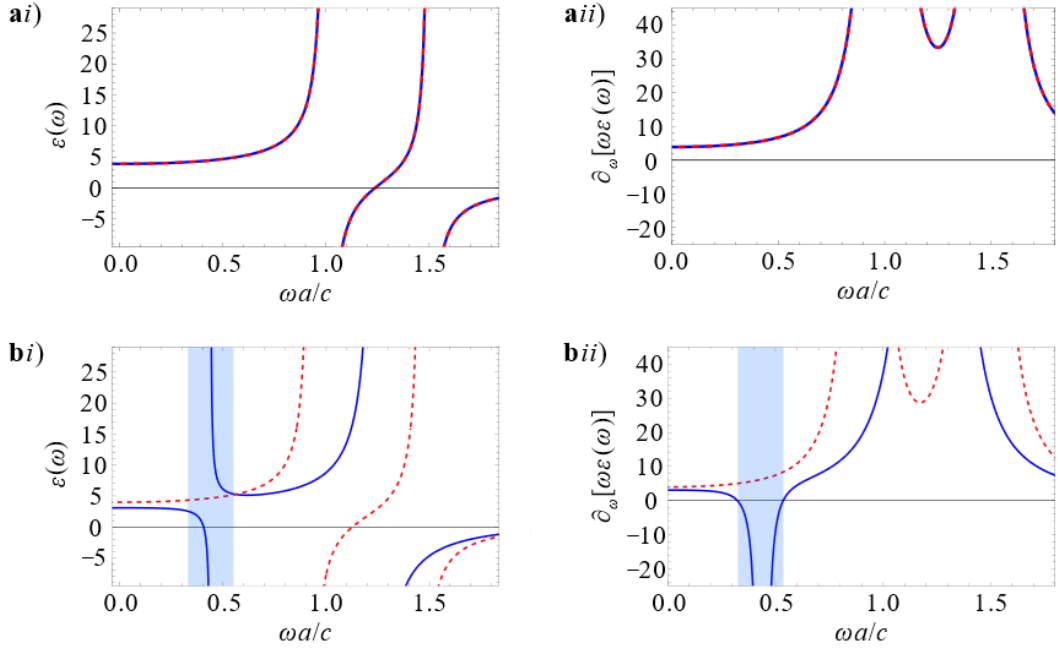

**Fig. 2** Effective permittivity i) $\varepsilon_{ef}(\omega)$ and ii) $\partial_{\omega}[\omega\varepsilon_{ef}(\omega)]$ as a function of the frequency $\omega$, for a binary crystal with $a_1 = a_2 = a/2$, $\omega_{p,1} = \omega_{p,2} = 2.0c/a$, $\omega_{0,1} = 1.0c/a$, $\omega_{0,2} = 1.5c/a$ and for the modulation speed **(a)** $v = 0$ and **(b)** $v = 0.25c$. Blue lines: our theory. Dashed red lines: ND-homogenization.

The most notable difference between Figs. 2ai) and 2bi) is the monotonic behavior of the permittivity dispersion. In particular, the dispersion predicted by the source-driven homogenization exhibits an anomalous (non-Foster) behavior such that in some spectral regions the permittivity *decreases* with the frequency. For transparent (i.e., weakly absorptive) passive linear materials, such a behavior is forbidden by the Kramers-Kronig relations [53]. In fact, the permittivity of a regular transparent material must increase monotonically with frequency due to well known analytical properties of Herglotz-Nevanlinna functions. On the other hand, spacetime modulated platforms are active, and thereby are not bound by the same constraints as passive materials. For example, in an active system it is possible to have $\mathrm{Im}\{\varepsilon(\omega)\} < 0$ in the upper-half frequency plane, different from passive systems [54]. Our results show that the active response of a dispersive spacetime crystal may be used to engineer an effective medium with an anomalous dispersion. As discussed in the Introduction, non-Foster materials are useful to compensate the "positive" dispersion of conventional Foster elements, enabling the design of broadband microwave and optical components, such as waveguides [42], antennas [43], cloaking devices [44], and others.

The anomalous dispersion persists even for very small modulation speeds, but typically with narrower bandwidths. This can be seen in Fig. 3, which depicts the effective permittivity for progressively smaller values of $v$. The spectral windows (with positive frequency) with anomalous permittivity dispersion are roughly centered at $-\langle \omega_0 \rangle + nv\frac{2\pi}{a}$ with $n$ a positive integer and $\langle \omega_0 \rangle$ some "effective" resonance frequency in the interval determined by $\omega_{0,1}$ and $\omega_{0,2}$. For small velocities, $\langle \omega_0 \rangle$ is approximately the spatial average of $\omega_0(x')$. The number of resonances increases for smaller values of $v$.

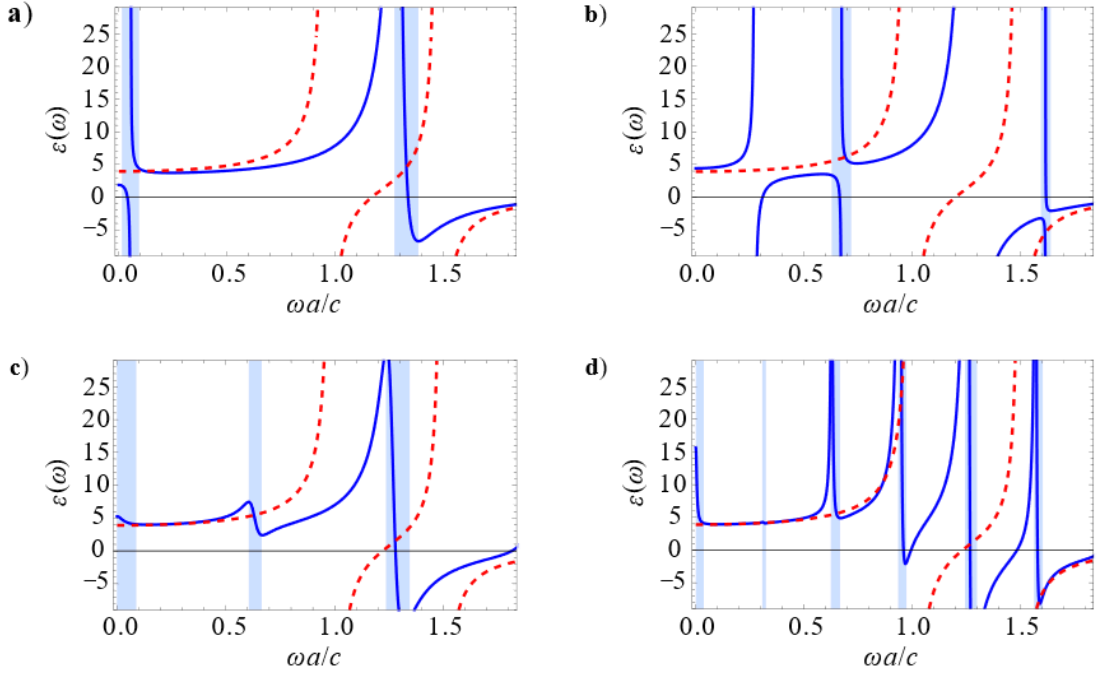

**Fig. 3** Effective permittivity of a binary crystal with the same parameters as in Fig. 2 for different modulation velocities: **(a)** $v = 0.20c$, **(b)** $v = 0.15c$, **(c)** $v = 0.10c$, **(d)** $v = 0.05c$. The spectral regions that exhibit an anomalous dispersion are highlighted by the vertical shaded blue strips. Blue lines: our theory. Dashed red lines: ND-homogenization.

The permittivity resonances may be attributed to the resonances of the materials at $\omega = \pm \omega_{0,n}$, which lead to resonances at $\omega' \approx \pm \langle \omega_0 \rangle - kv + m \frac{2\pi}{a} v$, with $m$ an integer, in the co-moving frame. This result is a consequence of the synthetic Doppler shift (term $-kv$) and of the band folding induced by the periodicity of the system in the co-moving frame. When such resonances are brought back to the laboratory frame, they emerge at $\omega \approx \pm \langle \omega_0 \rangle + m \frac{2\pi}{a} v$. As mentioned above, the positive frequency resonances associated with the anomalous dispersion are of the form $-\langle \omega_0 \rangle + nv \frac{2\pi}{a}$, i.e., are due to the band-folding of the negative frequency resonance. In other words, the spacetime folding of the dispersion of the bulk material creates "copies" of a negative (positive) resonance at $-\langle \omega_0 \rangle$ ($+\langle \omega_0 \rangle$) in the positive (negative) half of the frequency axis, which are responsible for the regions of anomalous dispersion. Thus, the non-Foster behavior results from the interaction of positive (negative) frequencies with negative (positive) resonances, which is only possible due to the combination of dispersion and spacetime band folding.

It is interesting to compare our dynamical homogenization with the theory of Ref. [32] for nondispersive materials. For a bi-layer non-magnetic and *non-dispersive* crystal the relative effective permittivity satisfies [32]:

$$\varepsilon_{\text{ef}}^{\text{ND}} = \frac{\varepsilon'}{\left(1 - \xi' v / c\right)^2 - \varepsilon' \mu' v^2 / c^2} \tag{24a}$$

with

$$\varepsilon' = f_1 \frac{\varepsilon_1}{1 - \varepsilon_1 v^2 / c^2} + f_2 \frac{\varepsilon_2}{1 - \varepsilon_2 v^2 / c^2}, \tag{24b}$$

$$\mu' = f_1 \frac{1}{1 - \varepsilon_1 v^2 / c^2} + f_2 \frac{1}{1 - \varepsilon_2 v^2 / c^2}, \tag{24c}$$

$$\xi' = -\frac{v}{c} \left[ f_1 \frac{\varepsilon_1}{1 - \varepsilon_1 v^2 / c^2} + f_2 \frac{\varepsilon_2}{1 - \varepsilon_2 v^2 / c^2} \right]. \tag{24d}$$

The effective permittivity $\varepsilon_{\text{ef}}^{\text{ND}}$ was derived under the assumption that the material parameters $\varepsilon_1$, $\varepsilon_2$ are independent of frequency [32]. It is natural to ask if $\varepsilon_{\text{ef}}^{\text{ND}}$ can be used to predict the effective permittivity of the dispersive crystal simply by replacing the symbols $\varepsilon_1$, $\varepsilon_2$ in the above formulas by the dispersive material permittivities $\varepsilon_1(\omega), \varepsilon_2(\omega)$, defined as in Eq. (23). We shall refer to such an approximate theory as non-dispersive (ND) homogenization.

The dashed red lines in Figs. 2ai), 2bi) and Fig. 3 were calculated with $\varepsilon_{\text{ef}}^{\text{ND}}$, using the procedure outlined in the previous paragraph. Although the ND-homogenization agrees precisely with the dynamical homogenization for $v=0$ (Fig. 2ai), the two methods predict totally different results when the modulation speed is nontrivial, particularly for moderate and large values of $\omega a/c$. In particular, the ND-homogenization misses completely the resonances associated with the spacetime band folding.

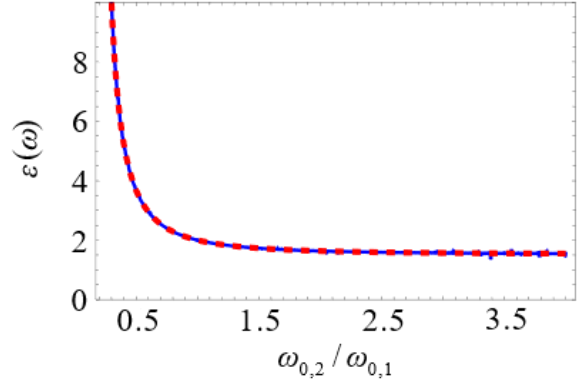

**Fig. 4** Effective permittivity of a binary crystal subject to a travelling-wave modulation with $v=0.25c$ for $\omega_{0,1} = \omega_{p,1} = \omega_{p,2} = 1000.0c/a$ and $\omega = 0.1c/a$ as a function of $\omega_{0,2}/\omega_{0,1}$. Blue lines: our theory. Dashed red lines: ND-homogenization.

Interestingly, as shown in Fig. 4, the ND-homogenization and the source-driven homogenization agree precisely in the limit $\omega_{p,n}^2 \to \infty, \omega_{0,n}^2 \to \infty$ with $\omega_{p,n}^2 / \omega_{0,n}^2 \to const. \equiv \varepsilon_n - 1$, i.e., in the non-dispersive limit (here, we neglect spurious resonances due to the spacetime band folding whose bandwidths are too narrow to represent graphically). Thus, our theory generalizes the results of previous work [32].

B.  *Negative stored energy*

As is well known, the energy stored in passive isotropic dielectric material in a transparency window is determined by (*V* represents the volume of the material) [53]:

$$\mathcal{E}_{av} = \frac{1}{4}\int_V dV \left[ \varepsilon_0 \frac{\partial}{\partial \omega}[\omega\varepsilon(\omega)]\mathbf{E}_\omega^*(\mathbf{r})\cdot\mathbf{E}_\omega(\mathbf{r}) + \mu_0 \mathbf{H}_\omega^*(\mathbf{r})\cdot\mathbf{H}_\omega(\mathbf{r}) \right]. \quad (25)$$

A time harmonic variation of the fields is implicit. The stored energy is the energy transferred to the material during the transient process that setups the stationary state field distribution. Thus, for a passive (weakly dissipative) material it is mandatory that $\frac{\partial}{\partial\omega}[\omega\varepsilon(\omega)] > 0$. This ensures that the energy transferred by the generator to the material is "positive", i.e., that it is necessary to pump energy into the material to reach the relevant steady state. In agreement with this property, the effective permittivity that characterizes the considered photonic crystals in the absence of a spacetime modulation satisfies $\frac{\partial}{\partial\omega}[\omega\varepsilon(\omega)] > 0$. This property is illustrated in Fig. 2aii) for a particular realization of a photonic crystal.

In Appendix C, it is demonstrated that $\mathcal{E}_{av}$ defined as in Eq. (25) may still have a physical meaning when the material is active, e.g., a homogenized spacetime crystal with $v \neq 0$. Specifically, provided $\varepsilon_{ef}(\omega)$ is real-valued for real-valued frequencies, then $\mathcal{E}_{av}$ determines the energy transferred from the external excitation (e.g., an antenna) during the transient period that precedes the steady-state regime. As discussed above, for conventional passive materials $\mathcal{E}_{av}$ must be strictly positive, and accordingly one needs that $\frac{\partial}{\partial\omega}[\omega\varepsilon(\omega)] > 0$. Differently, for an active system the direction of flow of energy from the generator to the material can be arbitrary due to the gainy nature of the material response. Hence, it is in principle possible to have situations in which $\frac{\partial}{\partial \omega}[\omega \varepsilon(\omega)] < 0$ and where $\mathcal{E}_{av} < 0$. Indeed, as shown in Fig. 2bii), our simulations (blue line) confirm that there are spectral windows wherein $\frac{\partial}{\partial \omega}[\omega \varepsilon(\omega)] < 0$, such that the "stored" energy is *negative*. This means that the material radiates away more energy than what it receives from the excitation in the transient process that leads to a steady state. As seen in Fig. 2bii), the ND-homogenization completely fails to predict such regimes. As might be expected, the spectral region wherein the stored energy is negative is approximately coincident with the region of anomalous permittivity dispersion.

## C. *Complex effective permittivity*

Notwithstanding the active nature of time-variant systems, in the previous examples the effective permittivity was always real-valued for real-valued frequencies. As discussed in Sect. III.C, the reality of $\varepsilon_{ef}$ is guaranteed when either (i) $\hat{L}_{co}$ is Hermitian or (ii) $\hat{L}_{co}$ is parity-time symmetric (see Appendix A).

Thus, in order that the non-Hermitian character of the spacetime crystal may manifest itself in the stationary-state effective response it is necessary to break both the PT-symmetry and the Hermitian property of $\hat{L}_{co}$. The operator $\hat{L}_{co}$ in Eq. (18b) is always PT-symmetric for binary crystals. Thereby, $\varepsilon_{ef}$ can be complex-valued in the real-frequency axis only if the unit cell of the crystal is formed by three or more layers (see Fig. 5a). In addition, we need to ensure that $\hat{L}_{co}$ is not Hermitian, which requires that $\omega_p(x') \neq const.$ (see Appendix A). Hence, next we consider ternary crystals such that $\omega_{p,1} \neq \omega_{p,2} \neq \omega_{p,3}$.

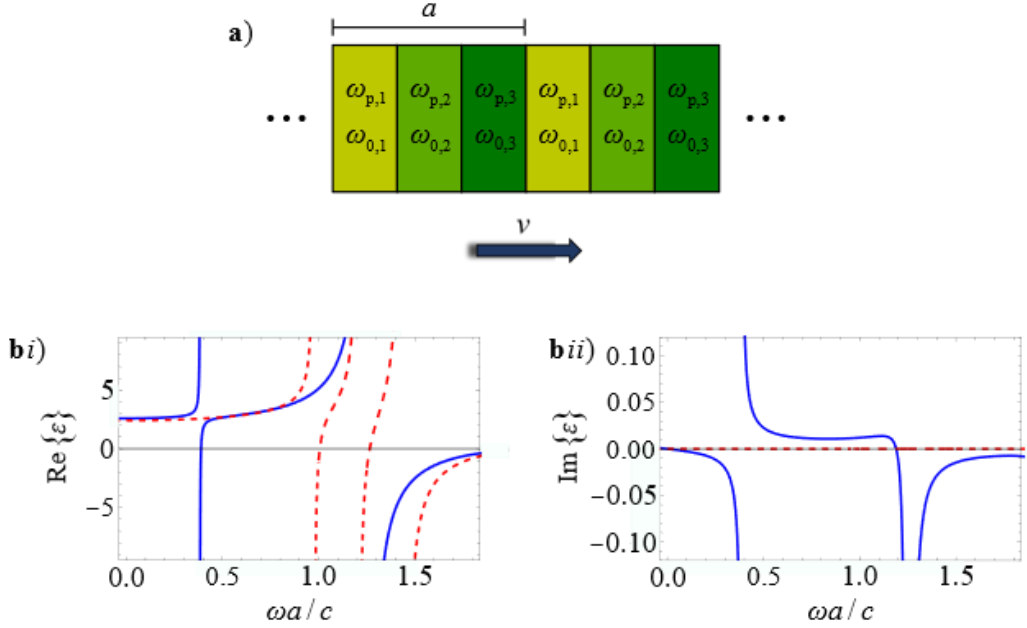

**Fig. 5** **(a)** Geometry of a dispersive spacetime crystal with a ternary structure. **(b)** Real and imaginary parts of the effective permittivity as a function of $\omega$ for $a_1 = a_2 = a_3 = a/3$, $\omega_{p,1} = 1.0c/a$, $\omega_{p,2} = 2.0c/a$, $\omega_{p,3} = 1.5c/a$, $\omega_{0,1} = 1.0c/a$, $\omega_{0,2} = 1.5c/a$, $\omega_{0,3} = 1.25c/a$, $v = 0.25c$. If $\text{Im}\{\varepsilon\} < 0$ ($\text{Im}\{\varepsilon\} > 0$), the material response is gainy (lossy). Solid blue lines: our theory. Dashed red lines: ND-homogenization.

As illustrated in Fig. 5b, consistent with the symmetry analysis of the previous paragraphs, such systems are generally characterized by a complex effective permittivity, such that the material response can alternate between gain ($\text{Im}\{\varepsilon\} < 0$) and loss ($\text{Im}\{\varepsilon\} > 0$) regimes depending on the frequency of operation. This type of non-Hermitian behavior persists for arbitrarily small velocities (not shown). In the example of Fig. 5, the source-driven and the ND-homogenization methods yield the same results in the static limit but differ substantially for $\omega a / c > 0.3$. In particular, the ND-homogenization (dashed red curves) fails to predict the existence of gain and loss regimes. It is worth noting that, while $\omega_0$ is modulated in Fig. 5, the non-Hermitian character of the operator $\hat{L}_{co}$ is independent of such modulation. In fact, it is still possible to obtain a complex effective permittivity with a constant parameter $\omega_0$.

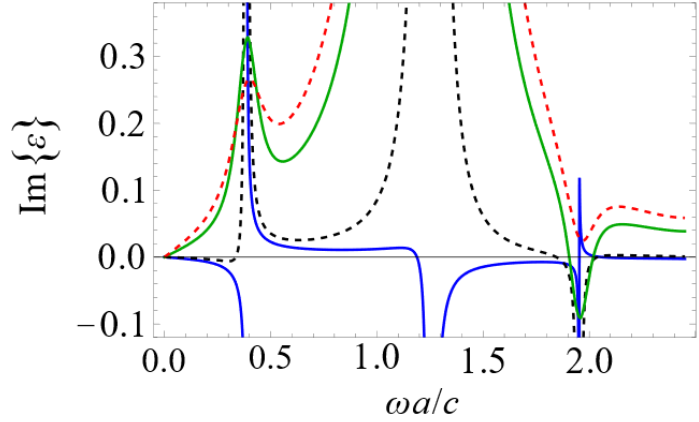

**Fig. 6** Imaginary part of the effective permittivity of a ternary spacetime crystal with the same parameters as Fig. 5 for different values of the collision frequency: $\Gamma=0.00 c/a$ (solid blue curve), $\Gamma=0.01 c/a$ (dashed black curve), $\Gamma=0.12 c/a$ (solid green curve), $\Gamma=0.18 c/a$ (dashed red curve).

It is possible to suppress the gainy response of the crystal by including a dissipative term in Eq. (13) as follows:

$$\left[\frac{\partial^2}{\partial t^2}+\Gamma\frac{\partial}{\partial t}+\omega_0^2(x-vt)\right]P(x,t)=\varepsilon_0\omega_p^2(x-vt)E(x,t). \qquad (26)$$

Here, $\Gamma$ represents a collision frequency, which for simplicity is assumed independent of space and time. In such a case the $\hat{L}_{co}$ operator in the co-moving frame must be modified as (still assuming that $k\approx 0$):

$$\hat{L}_{co}\to\hat{L}_{co,0}+i\frac{\Gamma}{\omega_p^2(x')}\left(-\omega+vi\frac{\partial}{\partial x'}\right), \qquad (27)$$

where $\hat{L}_{\text{co},0}$ is the operator defined in Eq. (18b). The effective permittivity can be numerically determined using the same procedure as in Sect. III.B. Figure 6 shows the imaginary component of the effective permittivity for different values of the collision frequency. As seen, for non-zero collision frequencies the gainy response is weakened, and for a sufficiently large collision frequency it can even be eliminated. It should be noted that for $\Gamma \approx 0$ the system response may have resonances nearby the real-frequency axis, some of which may possibly lie in the upper-half frequency plane (UHP). In fact, time-modulated systems are often characterized by parametric instabilities. Such resonances are expected to migrate to the lower-half frequency plane for a sufficiently large damping rate ($\Gamma > \Gamma_c$). Thus, the qualitative difference between the results with $\Gamma \approx 0$ and curves with large $\Gamma$ can be justified noting that, in the former case, the material resonances lie nearby the real-frequency axis, whereas in the latter case they are well below the real-frequency axis.

## V. Photonic band structure

In order to validate our theory, in the following we compare the exact photonic band structure of the spacetime crystals with the dispersion predicted by the homogenization.

Figure 7 shows the exact (blue lines) and approximate (dashed-black lines) band diagrams calculated for binary crystals with different parameters $\omega_{p,n}$, $\omega_{0,n}$ and $v$. The exact band structure of the dispersive spacetime crystals is determined using a transfer matrix approach detailed in Appendix D. On the other hand, the band structure predicted by our source-driven homogenization is found using the secular equation:

$$k^2 = \left(\frac{\omega}{c}\right)^2 \varepsilon_{\text{ef}}(\omega). \tag{28}$$

Remarkably, the effective medium model captures faithfully the dispersion of most of the photonic bands, even for relatively large values of $\omega a/c$ and $ka$, where the homogenization is expected to break down. In fact, for relatively small modulation speeds [Fig. 7a - 7c], i.e., $v^2/c^2 \ll 1$, the theory can model the dispersion of bands for regions that are considerably far away from the long-wavelength regime. For larger velocities [Fig. 7d], the effective medium becomes less accurate, as it predicts precisely only the dispersion of the 1$^{\text{st}}$ band.

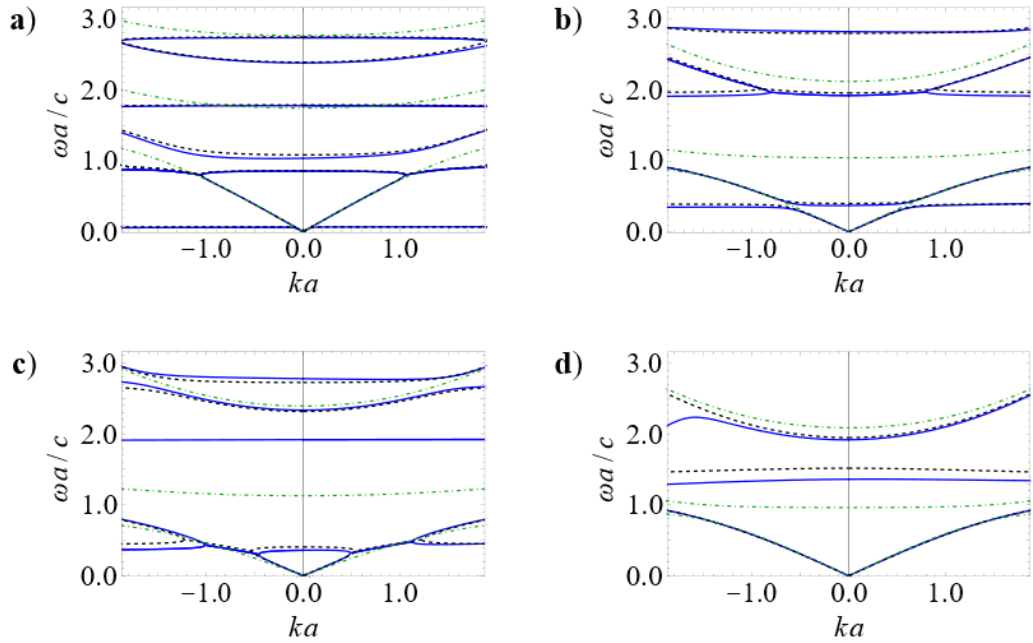

**Fig. 7** Band structure for binary crystals with $a_1 = a_2 = a/2$ and characterized by different parameters: **(a)** $\omega_{p,1} = \omega_{p,2} = 1.5c/a$, $\omega_{0,1} = 1.5c/a$, $\omega_{0,2} = 2.5c/a$, $v = 0.15c$; **(b)** $\omega_{p,1} = 1.0c/a$, $\omega_{p,2} = 2.0c/a$, $\omega_{0,1} = 1.0c/a$, $\omega_{0,2} = 1.5c/a$, $v = 0.25c$; **(c)** $\omega_{p,1} = \omega_{p,2} = 2.0c/a$. $\omega_{0,1} = 1.0c/a$, $\omega_{0,2} = 1.5c/a$,

$v = 0.25c$; **(d)** same as b) but with $v = 0.40c$. Blue lines: exact formulation. Dashed black lines: source-driven homogenization. Dash-dotted green lines: ND-homogenization.

We have also computed the band structure predicted by the ND-homogenization discussed in Sect. III (dash-dotted green lines). As shown in Fig. 7, typically the ND-homogenization model describes correctly the dispersion of the 1$^{st}$ photonic band in the static limit but misses completely the complex interactions resulting from the spacetime modulations at finite frequencies. Consequently, the ND-homogenization is unable to capture the quasi-flat bands that populate the spectrum of the spacetime crystal. For example, the ND-homogenization results for the spacetime crystals (b) and (d), which differ only by the value of the modulation velocity, are almost identical.

The comparison between the band structures of the exact and the homogenized models is designed to validate the effective material response to Bloch waves with a real-valued wavenumber $k$. Other methods can be used to evaluate the accuracy of our source-driven homogenization theory for complex-valued wavenumbers, such as assessing the response to a confined source distribution. However, these alternative analyses employ complex numerical tools and, for this reason, they lie outside the scope of this work.

The dispersion relation (28) shows that there is a direct link between the frequencies $\omega(k=0)$ of the band diagram and the zeroes of the effective permittivity. By comparing the band diagram in Fig. 7c with the corresponding effective permittivity in Fig. 2bi, it can be checked that the anomalous dispersion is associated with the second band. Since the ND-homogenization misses this band, it is also unable to predict this exotic behavior.

The band diagrams of Fig. 7 only depict the part of the spectrum with $ka < 2$. It is instructive to show the full band structure of a generic spacetime crystal. This is done in Fig. 8 for a crystal with the same parameters as in Fig. 7b. The imaginary part of $\omega$ due to the non-Hermitian effects is ignored for simplicity. The figure represents both the positive and negative frequency parts of the spectrum, and depicts the band structure both in the laboratory frame (panel *a*) and in the co-moving frame (panel *b*) (see Appendix D). To better illustrate the periodicity of the band structure, we represent three complete Brillouin zones.

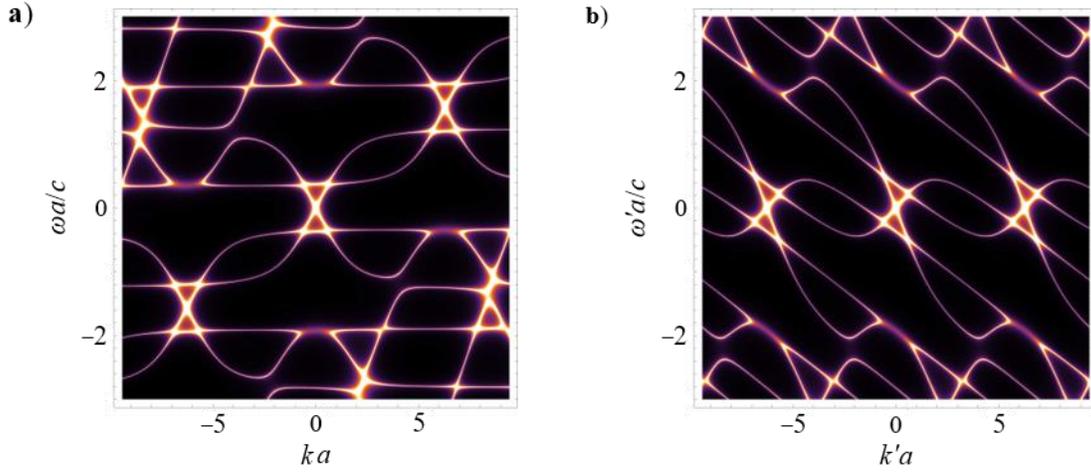

**Fig. 8** Exact band structure (bright curves) of a binary crystal with $\omega_{p,1} = 1.0c/a$, $\omega_{p,2} = 2.0c/a$, $\omega_{0,1} = 1.0c/a$, $\omega_{0,2} = 1.5c/a$ for the modulation speed $v = 0.25c$. **(a)** Laboratory frame. **(b)** Co-moving frame.

As seen in Fig. 8a, for large values of $k$ there is an evident asymmetry in the dispersion relation, so that $\omega(k) \neq \omega(-k)$, consistent with the nonreciprocity of time-variant systems. This spectral asymmetry is not captured by the (quasi-static) homogenization [see Fig. 7b] because $\varepsilon_{ef}(\omega)$ does not model the spatial dispersion effects. In contrast, the band diagram in the co-moving frame [Fig. 8b] is highly-asymmetric due to the synthetic Doppler shift effect (see Appendix D). Note that in the co-moving frame the band structure is periodic along the *k*-direction. In the lab-frame [Fig. 8a], the band structure periodicity is along an oblique direction of the $\omega - k$ plane determined by the synthetic Doppler shift.

The tilted lines in Fig. 8b represent the co-moving frame resonances at $\omega' \approx \pm \langle \omega_0 \rangle - kv + m\frac{2\pi}{a}v$, with *m* an integer, which were already discussed in Sect. IV.A. When such resonances are brought back to the laboratory frame, they emerge at $\omega \approx \pm \langle \omega_0 \rangle + m\frac{2\pi}{a}v$ leading to the quasi-flat bands. This creates an elaborate band structure where the modulation velocity *v* plays a crucial role in determining the location of these quasi-flat bands. For example, in Fig. 8a the first quasi-flat band occurs at $\omega \approx 0.3c/a$ (this quasi-flat band is also represented in Fig. 7b). The same mechanism is responsible for the quasi-flat band near the horizontal axis of Fig. 7a. It is underlined that these quasi flat-bands are correctly predicted by the homogenization.

In contrast, it may be shown that the quasi-flat band at $\omega \approx 1.9c/a$ in Fig. 7c describes a set of Bloch modes that oscillate very fast in space. Thus, the approximation of constant primed electromagnetic fields [Eq. (16)] is not applicable and the quasi-static homogenization is unable to predict that band.

## VI. Conclusions

We introduced a rigorous effective medium description of generic dispersive spacetime crystals. The formalism was applied to dispersive 1D-type spacetime crystals with a travelling-wave modulation. Interestingly, our theory reveals that such spacetime modulated systems can exhibit rather peculiar physics and extreme wave phenomena, such as an anomalous (non-Foster) dispersion with a negative stored energy density and, depending on the operating frequency, alternate between gain and loss regimes. All of these exotic properties are not captured by the homogenization theory for non-dispersive spacetime crystals presented in [32]. This demonstrates the importance of taking into account the dispersive behavior of the material response when developing an effective medium description. Furthermore, it was shown that the gainy regions and the anomalous dispersion regions arise due to negative frequency resonances of the bulk materials that are folded into the positive frequency spectrum as a result of the synthetic Doppler shift and of the periodicity of the system. The developed theory predicts rather accurately the photonic band diagram of the spacetime crystal, even for frequencies outside the usual scope of homogenization methods.

Our formalism can be readily extended to higher-dimensional photonic crystals and to arbitrary spacetime modulations. Thus, we expect that it can be a useful tool to describe the electrodynamics of generic spacetime crystals in the long wavelength limit.

**Acknowledgements:** This work was partially funded by the Institution of Engineering and Technology (IET), by the Simons Foundation (Award 733700) and by Instituto de Telecomunicações under Project No. UIDB/50008/2020.

## Appendix A: Conditions under which $\varepsilon_{\text{ef}}(\omega)$ is real-valued

In this Appendix, we investigate the conditions under which the effective permittivity $\varepsilon_{\text{ef}}(\omega)$ of the spacetime crystal is real-valued for a real-valued frequency. To begin with, we show that if the operator $\hat{L}_{co} = \hat{L}_{co}(\omega)$ defined by Eq. (18b) is Hermitian then $\varepsilon_{ef}(\omega)$ is real-valued.

To this end, it is useful to introduce the Green's function $G_\omega(x', x'_0)$ defined as the solution of

$$\hat{L}_{co} \cdot G_\omega(x', x'_0) = \delta(x' - x'_0). \tag{A1}$$

Evidently, the function $G(x')$ introduced in Sect. III.B can be expressed in terms of $G_\omega(x', x'_0)$ as $G(x') = \int_{-a/2}^{a/2} G_\omega(x', x'_0) dx'_0$. Hence, the parameter $\chi'_{ef}(\omega)$ [Eq. (19)] that controls the effective permittivity [Eq. (20)] can be written in terms of the Green's function as:

$$\chi'_{ef}(\omega) = \frac{1}{a} \int_{-a/2}^{a/2} dx' \int_{-a/2}^{a/2} dx'_0 G_\omega(x', x'_0). \tag{A2}$$

To proceed further, we note that from Eq. (A1) $\int_{-a/2}^{a/2} dx' G^*_\omega(x', x''_0) \hat{L}_{co} \cdot G_\omega(x', x'_0) = G^*_\omega(x'_0, x''_0)$. If the operator $\hat{L}_{co} = \hat{L}_{co}(\omega)$ is Hermitian, one can also write

$$\int_{-a/2}^{a/2} dx' G^*_\omega(x', x''_0) \hat{L}_{co} \cdot G_\omega(x', x'_0) = \int_{-a/2}^{a/2} dx' \left[ \hat{L}_{co} G_\omega(x', x''_0) \right]^* \cdot G_\omega(x', x'_0) = G_\omega(x''_0, x'_0). \tag{A3}$$

Thus, it follows that for a Hermitian operator:

$$G_\omega(x', x'_0) = G^*_\omega(x'_0, x') \qquad \text{(Hermitian operator)}. \tag{A4}$$

Using this result in Eq. (A2), one readily sees that:

$$[\chi'_{\text{ef}}(\omega)]^* = \frac{1}{a}\int_{-a/2}^{a/2} dx'_0 \int_{-a/2}^{a/2} dx' G_\omega(x'_0, x') = \chi'_{\text{ef}}(\omega). \tag{A5}$$

In these circumstances, the effective permittivity $\varepsilon_{\text{ef}}(\omega)$ also satisfies $\varepsilon_{\text{ef}}(\omega) = [\varepsilon_{\text{ef}}(\omega)]^*$, i.e., it is real-valued.

A simple inspection of Eq. (18b) shows that for $\omega$ real-valued and $\omega_p(x') = const.$ the operator $\hat{L}_{\text{co}} = \hat{L}_{\text{co}}(\omega)$ is Hermitian. Thereby, in these conditions the effective permittivity is real-valued for a real-valued frequency. In contrast, when $\omega_p(x')$ is a nontrivial function, the operator is not Hermitian, and hence the permittivity may be complex-valued.

Interestingly, even when $\hat{L}_{\text{co}} = \hat{L}_{\text{co}}(\omega)$ is non-Hermitian it is possible to guarantee that $\varepsilon_{\text{ef}}(\omega)$ is real-valued for real-valued frequencies if the system is parity-time (PT) symmetric [49]. The invariance under PT symmetry corresponds to the condition $\hat{L}_{\text{co}}(x', i\partial_{x'}, \omega) = \hat{L}_{\text{co}}(-x', -i\partial_{x'}, \omega)^*$. For a PT symmetric system, the Green's function satisfies:

$$G_\omega(x', x'_0) = G_\omega^*(-x', -x'_0) \quad \text{(PT-symmetric operator)}. \tag{A6}$$

In such case, one can write

$$[\chi'_{\text{ef}}(\omega)]^* = \frac{1}{a}\int_{-a/2}^{a/2} dx'_0 \int_{-a/2}^{a/2} dx' G_\omega(-x', -x'_0) = \chi'_{\text{ef}}(\omega). \tag{A7}$$

This confirms that the PT-symmetry of the operator ($\hat{L}_{\text{co}}(x', i\partial_{x'}, \omega) = \hat{L}_{\text{co}}(-x', -i\partial_{x'}, \omega)^*$) guarantees that the effective permittivity is real-valued.

# Appendix B: Solution of the homogenization problem

For a binary crystal with layers of the same length ($a_1 = a_2 = a/2$) the coefficients $A_1, A_2, B_1, B_2$ in Eq. (21) can be found imposing the boundary conditions discussed in the main text. This yields:

$$A_1 = \frac{C_1 - C_2}{D} e^{i\alpha_1 \frac{a}{4}} \alpha_2 \beta_2 \left( e^{i\beta_1 \frac{a}{2}} - 1 \right) \left( e^{i\alpha_2 \frac{a}{2}} - e^{i\beta_2 \frac{a}{2}} \right) +$$
$$\frac{C_1 - C_2}{D} e^{i\alpha_1 \frac{a}{4}} \left[ \alpha_2 \beta_1 \left( e^{i(\alpha_2 + \beta_1)\frac{a}{2}} - 1 \right) \left( e^{i\beta_2 \frac{a}{2}} - 1 \right) - \beta_1 \beta_2 \left( e^{i\alpha_2 \frac{a}{2}} - 1 \right) \left( e^{i(\beta_1 + \beta_2)\frac{a}{2}} - 1 \right) \right]$$
(B1a)

$$B_1 = \frac{C_2 - C_1}{D} e^{i\beta_1 \frac{a}{4}} \alpha_2 \beta_2 \left( e^{i\alpha_1 \frac{a}{2}} - 1 \right) \left( e^{i\alpha_2 \frac{a}{2}} - e^{i\beta_2 \frac{a}{2}} \right)$$
$$+ \frac{C_2 - C_1}{D} e^{i\beta_1 \frac{a}{4}} \left[ \alpha_1 \alpha_2 \left( e^{i(\alpha_1 + \alpha_2)\frac{a}{2}} - 1 \right) \left( e^{i\beta_2 \frac{a}{2}} - 1 \right) - \alpha_1 \beta_2 \left( e^{i\alpha_2 \frac{a}{2}} - 1 \right) \left( e^{i(\alpha_1 + \beta_2)\frac{a}{2}} - 1 \right) \right]$$
(B1b)

$$A_2 = \frac{C_2 - C_1}{D} e^{i\alpha_2 \frac{a}{4}} \alpha_1 \beta_1 \left( e^{i\beta_2 \frac{a}{2}} - 1 \right) \left( e^{i\alpha_1 \frac{a}{2}} - e^{i\beta_1 \frac{a}{2}} \right) +$$
$$+ \frac{C_2 - C_1}{D} e^{i\alpha_2 \frac{a}{4}} \left[ \alpha_1 \beta_2 \left( e^{i(\alpha_1 + \beta_2)\frac{a}{2}} - 1 \right) \left( e^{i\beta_1 \frac{a}{2}} - 1 \right) - \beta_1 \beta_2 \left( e^{i\alpha_1 \frac{a}{2}} - 1 \right) \left( e^{i(\beta_1 + \beta_2)\frac{a}{2}} - 1 \right) \right]$$
(B1c)

$$B_2 = \frac{C_1 - C_2}{D} e^{i\beta_2 \frac{a}{4}} \alpha_1 \beta_1 \left( e^{i\alpha_2 \frac{a}{2}} - 1 \right) \left( e^{i\alpha_1 \frac{a}{2}} - e^{i\beta_1 \frac{a}{2}} \right) +$$
$$+ \frac{C_1 - C_2}{D} e^{i\beta_2 \frac{a}{4}} \left[ \alpha_1 \alpha_2 \left( e^{i(\alpha_1 + \alpha_2)\frac{a}{2}} - 1 \right) \left( e^{i\beta_1 \frac{a}{2}} - 1 \right) - \alpha_2 \beta_1 \left( e^{i\alpha_1 \frac{a}{2}} - 1 \right) \left( e^{i(\alpha_2 + \beta_1)\frac{a}{2}} - 1 \right) \right]$$
(B1d)

where $D$ is defined by:

$$D = (\alpha_2\beta_2 + \alpha_1\beta_1)\left(e^{i\alpha_1\frac{a}{2}} - e^{i\beta_1\frac{a}{2}}\right)\left(e^{i\alpha_2\frac{a}{2}} - e^{i\beta_2\frac{a}{2}}\right)$$

$$+ (\alpha_1\alpha_2 + \beta_1\beta_2)\left(e^{i(\alpha_1+\alpha_2)\frac{a}{2}} - 1\right)\left(e^{i(\beta_1+\beta_2)\frac{a}{2}} - 1\right) \quad \text{(B1e)}$$

$$- (\alpha_1\beta_2 + \alpha_2\beta_1)\left(e^{i(\alpha_2+\beta_1)\frac{a}{2}} - 1\right)\left(e^{i(\alpha_1+\beta_2)\frac{a}{2}} - 1\right)$$

## Appendix C: Energy Transferred to a Gainy Dispersive Medium in the Transient Period

Let us consider a dispersive effective medium characterized by the permittivity $\varepsilon(\omega)$. For the problem of interest $\varepsilon(\omega) = \varepsilon_{ef}(\omega)$ represents the effective permittivity of the spacetime crystal. Note that within the effective medium description the material response is homogeneous and time-invariant. However, since the spacetime crystal is active, the permittivity $\varepsilon(\omega)$ is not bound by the usual constraints that arise from the Kramers-Kronig relations. In this Appendix, it is supposed that $\varepsilon(\omega)$ is analytic in the upper-half frequency plane (UHP) (so that the material response is stable and causal), but it does not need to satisfy the constraint $\text{Im}\{\varepsilon(\omega)\} \geq 0$ in the UHP. Due to this reason, it is possible to have regimes with anomalous permittivity dispersion (see Fig. 2). Furthermore, it is assumed that $\varepsilon(\omega)$ is real-valued for real-valued frequencies, which is the case of homogenized spacetime crystals characterized by an operator $\hat{L}_{co}$ that is Hermitian.

Let us find the energy $\mathcal{E}$ transferred to the active material by some external excitation (e.g., an antenna) during the transient period that leads to a stationary state. In the stationary state the electric field is of the form $\mathbf{E}(\mathbf{r},t) = \text{Re}\{\mathbf{E}_\omega(\mathbf{r})e^{-i\omega t}\}$ with $\omega$ real-valued. The external excitation is modeled by an electric current density $\mathbf{j}_{\text{ext}}$. For simplicity, we imagine that the system is surrounded by closed boundaries (e.g., metallic walls).

From the Poynting theorem, it is known that $\nabla \cdot \mathbf{S} = -p - p_{\text{ext}}$ where $\mathbf{S}(\mathbf{r},t) = \mathbf{E} \times \mathbf{H}$ is the Poynting vector, $p(\mathbf{r},t) = \mathbf{E} \cdot \frac{\partial \mathbf{D}}{\partial t} + \mathbf{H} \cdot \frac{\partial \mathbf{B}}{\partial t}$ and $p_{\text{ext}}(\mathbf{r},t) = \mathbf{E} \cdot \mathbf{j}_{\text{ext}}$. The quantity $p_{\text{ext}}$ represents the instantaneous power per unit of volume supplied by the excitation to the system. It is supposed that at initial time $t = -\infty$ the fields vanish. We wish to find the energy $\mathcal{E}(t) = -\int_V dV \int_{-\infty}^{t} dt\, p_{\text{ext}}(\mathbf{r},t)$ transferred from the generator to the material during the transient period. Note that as the material is active it may also be internally excited by other internal mechanisms that determine the gainy type response.

Integrating $\nabla \cdot \mathbf{S} = -p - p_{\text{ext}}$ over the volume of the material and taking into account that the boundaries are closed one finds that:

$$\mathcal{E}(t) = \int_{-\infty}^{t} dt \int_V p(\mathbf{r},t) dV. \tag{C1}$$

The external excitation drives the fields, and thereby controls how the fields change in time. It is supposed that the time-harmonic excitation is turned on slowly, which corresponds to a time variation of the type $e^{-i\omega t} = e^{-i\omega' t} e^{\omega'' t}$ with $\omega = \omega' + i\omega''$ and $\omega'' > 0$ very small. Notice that $\lim_{t \to -\infty} e^{-i\omega t} = 0$. We suppose that the field time-variation is inherited from the time variation of the excitation, which is certainly valid if the material

-37-

sample is small enough so that retardation effects are of secondary importance. In the outlined conditions one can assume that $\mathbf{E}(\mathbf{r},t) = \text{Re}\{\mathbf{E}_\omega(\mathbf{r})e^{-i\omega t}\}$ and $\mathbf{D}(\mathbf{r},t) = \text{Re}\{\mathbf{D}_\omega(\mathbf{r})e^{-i\omega t}\}$, etc so that:

$$\mathcal{E}(t) = \int_V dV \int_{-\infty}^t dt\, \frac{e^{2\omega''t}}{2} \text{Re}\{-i\omega \mathbf{E}_\omega(\mathbf{r}) \cdot \mathbf{D}_\omega(\mathbf{r}) e^{-i2\omega't} - i\omega \mathbf{E}_\omega^*(\mathbf{r}) \cdot \mathbf{D}_\omega(\mathbf{r})\}$$
$$= \int_V dV \int_{-\infty}^t dt\, \frac{e^{2\omega''t}}{2} \text{Re}\{-i\omega \mathbf{H}_\omega(\mathbf{r}) \cdot \mathbf{B}_\omega(\mathbf{r}) e^{-i2\omega't} - i\omega \mathbf{H}_\omega^*(\mathbf{r}) \cdot \mathbf{B}_\omega(\mathbf{r})\}$$
(C2)

Integrating explicitly in time:

$$\mathcal{E}(t) = \frac{e^{2\omega''t}}{4} \int_V dV\, \text{Re}\left\{\frac{-i\omega}{\omega'' - i\omega'} \mathbf{E}_\omega(\mathbf{r}) \cdot \mathbf{D}_\omega(\mathbf{r}) e^{-i2\omega't} - \frac{i\omega}{\omega''} \mathbf{E}_\omega^*(\mathbf{r}) \cdot \mathbf{D}_\omega(\mathbf{r})\right\}$$
$$+ \frac{e^{2\omega''t}}{4} \int_V dV\, \text{Re}\left\{\frac{-i\omega}{\omega'' - i\omega'} \mathbf{H}_\omega(\mathbf{r}) \cdot \mathbf{B}_\omega(\mathbf{r}) e^{-i2\omega't} - \frac{i\omega}{\omega''} \mathbf{H}_\omega^*(\mathbf{r}) \cdot \mathbf{B}_\omega(\mathbf{r})\right\}$$
(C3)

In the next step we use $\mathbf{D}_\omega = \varepsilon_0 \varepsilon(\omega) \mathbf{E}_\omega$ and $\mathbf{B}_\omega = \mu_0 \mathbf{H}_\omega$. In order to reach a steady state we let $\omega'' \to 0^+$. The terms proportional to $e^{-i2\omega't}$ have no singularities in the limit $\omega'' \to 0^+$. They are oscillatory terms with frequency $2\omega'$ and on average (in a cycle with time period $T = \frac{2\pi}{\omega'}$) vanish. Hence, they are dropped:

$$\mathcal{E}_{av} = \lim_{\omega'' \to 0^+} \frac{1}{4} \int_V dV \left[ \text{Re}\left\{\frac{-i\omega\varepsilon(\omega)}{\omega''}\right\} \varepsilon_0 \mathbf{E}_\omega^*(\mathbf{r}) \cdot \mathbf{E}_\omega(\mathbf{r}) + \text{Re}\left\{\frac{-i\omega}{\omega''}\right\} \mu_0 \mathbf{H}_\omega^*(\mathbf{r}) \cdot \mathbf{H}_\omega(\mathbf{r}) \right].$$ (C4)

The outstanding terms are seemingly singular due to the factor $1/\omega''$. The singularity can be handled using $\omega = \omega' + i\omega''$ and the Taylor expansion $\omega\varepsilon(\omega) \approx \omega'\varepsilon(\omega') + i\omega'' \frac{\partial}{\partial \omega}[\omega\varepsilon(\omega)]_{\omega = \omega'}$. Taking into account that $\omega'\varepsilon(\omega')$ is real-

-38-

valued, one finds that the energy transferred from the excitation to the material during the transient period is:

$$\mathcal{E}_{av} = \frac{1}{4}\int_V dV \left[ \varepsilon_0 \frac{\partial}{\partial \omega}[\omega\varepsilon(\omega)]_{\omega=\omega'} \mathbf{E}_\omega^*(\mathbf{r})\cdot\mathbf{E}_\omega(\mathbf{r}) + \mu_0 \mathbf{H}_\omega^*(\mathbf{r})\cdot\mathbf{H}_\omega(\mathbf{r}) \right]. \quad (C5)$$

Note that due to the reality of $\varepsilon(\omega')$ there is no net transfer of energy between the material and the excitation after the steady-state is reached.

## Appendix D: Dispersion of the Bloch waves

In this Appendix, we present the formalism used to find the dispersion of the Bloch waves of the dispersive spacetime crystal.

In a first step, we calculate the spectrum of the crystal in the co-moving frame. To this end, we solve Eq. (15) without any excitation ($j_{e,0} = 0$) and with $\partial/\partial t' = -i\omega'$. It is convenient to introduce the auxiliary variable $J = \frac{\partial}{\partial t}P = \left(\frac{\partial}{\partial t'} - v\frac{\partial}{\partial x'}\right)P$ to reduce the problem to a system of first-order differential equations in $x'$:

$$-i\frac{\partial E'}{\partial x'} = \mu_0 \omega' H, \qquad -i\frac{\partial H'}{\partial x'} = \omega'(\varepsilon_0 E + P), \quad (D1a)$$

$$\left(-i\omega' - v\frac{\partial}{\partial x'}\right)J + \omega_0^2(x')P = \varepsilon_0 \omega_p^2(x')E, \quad (D1b)$$

$$J = \left(-i\omega' - v\frac{\partial}{\partial x'}\right)P. \quad (D1c)$$

The primed fields are defined as in Eq. (16). In matrix notation, this system of equations can be rewritten as:

-39-

$$-i\frac{\partial}{\partial x'}\underbrace{\begin{pmatrix} E' \\ H' \\ P \\ J \end{pmatrix}}_{\psi(x')} = \underbrace{\begin{pmatrix} \frac{\omega'\mu_0\varepsilon_0 v}{1-v^2/c^2} & \frac{\omega'\mu_0}{1-v^2/c^2} & \frac{\omega'\mu_0 v}{1-v^2/c^2} & 0 \\ \frac{\omega'\varepsilon_0}{1-v^2/c^2} & \frac{\omega'\varepsilon_0\mu_0 v}{1-v^2/c^2} & \omega'\left(1+\frac{\varepsilon_0\mu_0 v^2}{1-v^2/c^2}\right) & 0 \\ 0 & 0 & -\frac{\omega'}{v} & \frac{i}{v} \\ \frac{1}{v}\frac{i\varepsilon_0\omega_{\mathrm{p}}^2(x')}{1-v^2/c^2} & \frac{i\varepsilon_0\mu_0\omega_{\mathrm{p}}^2(x')}{1-v^2/c^2} & -\frac{i\omega_0^2(x')}{v}+\frac{i\varepsilon_0\mu_0\omega_{\mathrm{p}}^2(x')v}{1-v^2/c^2} & -\frac{\omega'}{v} \end{pmatrix}}_{\mathbf{M}(x',\omega')}\begin{pmatrix} E' \\ H' \\ P \\ J \end{pmatrix} \quad \text{(D2)}$$

where we used Eq. (17) to express the unprimed fields in terms of the primed fields. The general solution of the problem can be written in terms of a propagator that links the state vector $\mathbf{\psi} = \begin{bmatrix} E' & H' & P & J \end{bmatrix}^T$ evaluated at different points of the crystal:

$$\mathbf{\psi}(x') = U_{\omega'}(x') \cdot \mathbf{\psi}(0). \quad \text{(D3)}$$

For a spacetime binary crystal, the propagator for points separated by a lattice period is $U_{\omega'}(a_1 + a_2) = e^{i\mathbf{M}_2(\omega')a_2} e^{i\mathbf{M}_1(\omega')a_1}$, with $\mathbf{M}_n(\omega')$ the value of the matrix $\mathbf{M}(x',\omega')$ at the $n$-th homogeneous layer ($n=1,2$). As the Bloch modes in the co-moving frame must satisfy $\mathbf{\psi}(a) = e^{ik'a}\mathbf{\psi}(0)$ it follows that the dispersion characteristic is given by:

$$\det\left(e^{ik'a}\mathbf{1} - U_{\omega'}(a)\right) = 0. \quad \text{(D4)}$$

In particular, one sees that for a given $\omega'$ the propagation constants $k'$ of the Bloch waves are such that $e^{ik'a}$ are the eigenvalues of the propagator $U$. Finally, to transform the band structure $\omega'(k')$ in the co-moving frame to the lab frame, we use the synthetic Doppler shift formula, i.e., $\omega(k) = \omega'(k) + vk$.